# APPLICATION OF SCANNING MID-IR-LASER MICROSCOPY FOR CHARACTERIZATION OF SEMICONDUCTOR MATERIALS FOR PHOTOVOLTAICS

V. P. Kalinushkin [†], O. V. Astafiev [‡,1], and V. A. Yuryev [‡]

[†] *General Physics Institute of RAS, 38 Vavilov Street, Moscow 117942, Russia.*
Tel.: +7 (095) 132 8126. Facsimile: +7 (095) 135 1330. E-mail: vkalin@kapella.gpi.ru.
[‡] *Natural Science Center of General Physics Institute of RAS.*
Tel.: +7 (095) 132 8144. Facsimile: +7 (095) 135 1330. E-mail: vyuryev@kapella.gpi.ru.

The scanning mid-IR-laser microscopy was previously demonstrated as an effective tool for characterization of different semiconductor crystals.[1–4] Now the technique has been successfully applied for the investigation of CZ $Si_xGe_{1-x}$ — a promising material for photovoltaics — and multicrystalline silicon for solar cells.

A pictorial diagram of the scanning mid-IR-laser microscope is given in Fig. 1. Presently, the instrument operates in two main modes: a mode of the scanning low-angle mid-IR light scattering (SLALS) — the basic regime of its operation — which is sensitive to aggregations of ionized point defects (or free carrier accumulations) and any inhomogeneities in the distribution of free carriers (some restrictions are imposed on the accumulation or inhomogeneity characteristic sizes and its profile function,[3] however) and a mode of the optical beam induced light scattering (OLALS) — an optical analog of EBIC or OBIC — which reveals recombination active defects and enables carrier lifetime mapping. Both modes are the dark-field ones. Besides, analogous regimes of the microscope operation are available in the bright field (mid-IR light absorption) sub-mode.

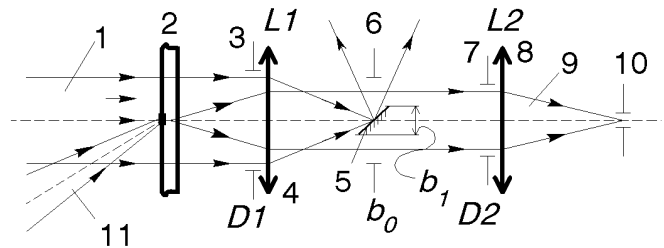

**Fig. 1.** An optical diagram of the scanning mid-IR-laser microscope: (1) the probe IR-laser beam (routinely, CO or $CO_2$-laser), (2) the studied sample (can be placed in a cryostat or a furnace) in the front focal plane of the lens L1, (3) an aperture with a diameter $D_1$ in the plane of the lens L1, (4) the lens L1, (5) an opaque screen or a mirror with a radius $b_1$ in the back focal plane of the lens L1, (6) a diaphragm with a radius $b_0$ in the back focal plane of the lens L1, (7) an aperture with a diameter $D_2$ in the plane of the lens L2, (8) the lens L2, (9) the scattered wave, (10) an IR photodetector in the back focal plane of the lens L2, (11) an exciting laser beam (used in the OLALS mode).

At present, low-temperature SLALS and OLALS facilities which would enable the defect composition analysis[3] as well as the magneto-optical SLALS are under the development. The defect profile option and quantitative carrier lifetime and concentration mapping will also be available soon.

SLALS and OLALS micrographs of the samples of single-crystalline CZ $Si_{1-x}Ge_x$ alloy with Ge content from 2.2 to 4.7 at. % are presented in Fig. 2. Two areas were revealed in the X-ray topographs of these crystals: the area free of striation and dislocations around the wafer centers (area I) and the area containing striation and dislocations in the periphery of the wafers (area II).[4] SLALS pictures shows the striation in the area II and no striation in the area I. OLALS pictures demonstrate that no or low (Fig. 2 (*h*)) recombination contrast is usually caused with the grown-in striations in the crystal bulk, although a high contrast was revealed in Fig. 2 (*l*). The second type of defects manifested as dark stripes in the OLALS micrographs (Fig. 2 (*b*),(*l*)) can likely be identified as dislocations and dislocation walls which are registered in X-ray patterns of the area II and revealed by etching. The last type of defects observed are those seen as black spots in the OLALS patterns (Fig. 2 (*b*),(*d*),(*f*),(*h*),(*j*)). They were present in both areas and appeared to have a non-dislocation origin: some non-dislocation defects found in both areas by the selective etching may be similar to the defects revealed by OLALS. The latter defects seem to be the main lifetime (and cell efficiency) killing extended defects in the studied material.

---

[1] Present address: Tokyo University, Department of Basic Science, Komaba 3-8-1, Meguro-ku, Tokyo, 153. E-mail: astf@mujin.c.u-tokyo.ac.jp.

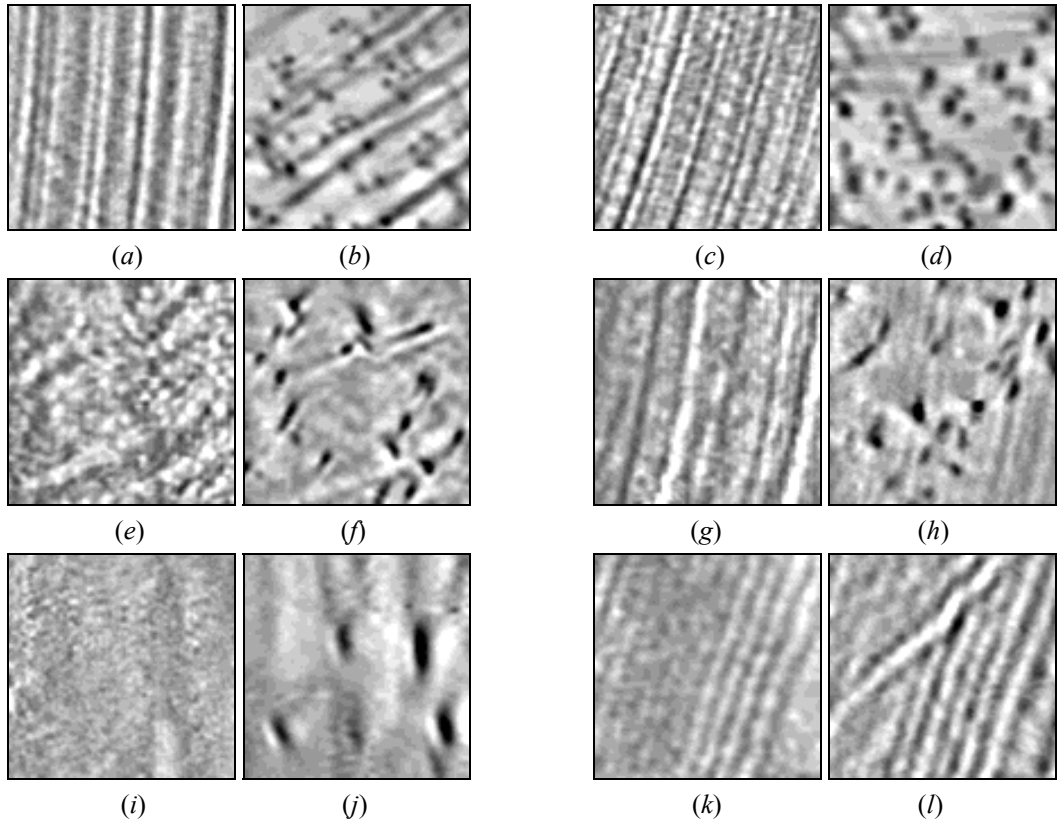

**Fig. 2** Couples of SLALS (*a*),(*c*),(*e*),(*g*),(*i*),(*k*) and OLALS (*b*),(*d*),(*f*),(*h*),(*j*),(*l*) micrographs of the same regions of $Si_{1-x}Ge_x$ wafers (1×1 mm); (*a*)–(*d*): p-type CZ Si (100), 4 at. % of Ge; (*e*)–(*h*): p-type CZ Si (111), 4.7 at. % of Ge; (*i*)–(*l*): n-type CZ Si (111), 2.2 at. % of Ge; (*e*),(*f*),(*i*),(*j*) are close to the wafer centers, the rest are far from the centers.

Fig. 3 shows the recombination contrast of grain boundaries in two samples of multicrystalline silicon for solar cells. It is clear that scanning mid-IR-laser microscopy can be effectively used for the investigation and testing of the efficiency of grain boundary passivation in this material.

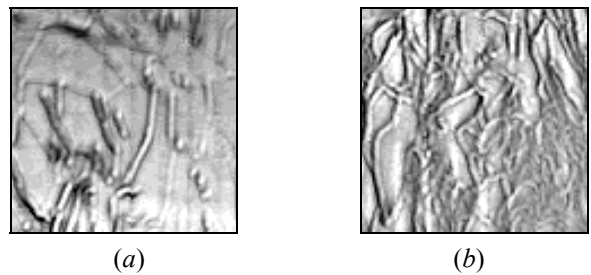

**Fig. 3** Micrographs of multicrystalline silicon for solar cells (the optical beam induced absorption sub-mode, 4×4 mm). The darker the image the shorter the lifetime is. Grain boundaries are clearly seen.

1. O. V. Astafiev, V. P. Kalinushkin, and V. A. Yuryev, *Inst. Phys. Conf. Ser.* **146** (1995) 775.
2. O. V. Astafiev, V. P. Kalinushkin, and V. A. Yuryev, *Inst. Phys. Conf. Ser.* **149** (1996) 361.
3. O. V. Astafiev, V. P. Kalinushkin, and V. A. Yuryev, *Rev. Sci. Instrum.* **70** (1999), in press.
4. O. V. Astafiev, V. P. Kalinushkin, and N. A. Abrosimov, *MRS Symp. Proc.* Vol. **442** (1997) 43.